\documentclass[letterpaper, 10 pt, conference]{ieeeconf}  

\IEEEoverridecommandlockouts                              

\usepackage{cite}
\usepackage{amsmath,amssymb,amsfonts}
\usepackage{algorithmic}
\usepackage{graphicx}
\usepackage{textcomp}
\usepackage{xcolor}
\usepackage[caption=false,font=footnotesize]{subfig}
\usepackage{booktabs}
\usepackage{multirow}
\usepackage{makecell}
\usepackage{url}

\title{\LARGE \bf
Phoneme-Level Deepfake Detection Across Emotional Conditions Using Self-Supervised Embeddings
}

\author{
    Vamshi Nallaguntla$^{1,4}$,
    Shruti Kshirsagar$^{1,5}$,
    and Anderson R. Avila$^{2,3,6}$%
    \thanks{$^{1}$Wichita State University, Wichita, KS, USA. }%
    \thanks{$^{2}$Institut national de la recherche scientifique (INRS--EMT), QC, Canada}%
    \thanks{$^{3}$INRS-UQO Mixed Research Unit on Cybersecurity, Gatineau, Canada}%
    \thanks{$^{4}$vxnallaguntla@shockers.wichita.edu, $^{5}$shruti.kshirsagar@wichita.edu}%
    \thanks{$^{6}$anderson.avila@inrs.ca}
}

\begin{document}

\maketitle
\thispagestyle{empty}
\pagestyle{empty}




\begin{abstract}
Recent advances in emotional voice conversion (EVC) have enabled the generation of expressive synthetic speech, raising new concerns in audio deepfake detection. Existing approaches treat speech as a homogeneous signal and largely overlook its internal phonetic structure, limiting their interpretability in emotionally conditioned settings. In this work, we propose a phoneme-level framework to analyze emotionally manipulated synthetic speech using real and EVC-generated speech under matched emotional conditions with shared transcripts, phoneme-aligned TextGrids, and WavLM-based embeddings. Our results show that phoneme behavior varies across categories, with complex vowels and fricatives exhibiting higher divergence while simpler phonemes remain more stable. Phonemes with larger distributional differences are also found to be more easily detected, consistently across multiple emotions and synthesis systems. These findings demonstrate that phoneme-level analysis is an effective and interpretable approach for detecting emotionally manipulated synthetic speech. \end{abstract}

Keywords- Audio deepfake detection, Emotional voice conversion, Phoneme-level analysis, WavLM, Kullback--Leibler divergence, Support vector machine, self-supervised learning, synthetic speech

\section{INTRODUCTION}
\label{introduction}

Recent advances in generative artificial intelligence have dramatically improved 
the quality and naturalness of synthetic speech, enabling systems that can 
produce human-like audio that is increasingly difficult to distinguish from 
genuine recordings \cite{ren2019fastspeech, prenger2019waveglow}. Among these 
developments, emotional voice conversion (EVC) has emerged as a particularly 
powerful capability, allowing systems to manipulate the emotional content of 
speech while preserving speaker identity and linguistic information 
\cite{kameoka2018stargan, qian2019autovc}. These technologies enable applications in human--computer interaction, virtual agents, assistive communication, and entertainment. However, they also pose serious risks: emotionally manipulated speech can deceive listeners, alter perceived intent, fabricate sentiment, and undermine the credibility of audio in high-stakes scenarios such as journalism, legal proceedings, and public discourse \cite{yi2023audio, delgado2021asvspoof}.

Detecting such audio deepfakes is fundamentally challenging because modern EVC 
systems produce speech that closely mimics the acoustic and prosodic 
characteristics of natural emotional expression. Existing deepfake detection 
approaches have primarily focused on utterance-level or frame-level 
representations derived from spectral features or self-supervised learning (SSL) 
models \cite{baevski2020wav2vec, chen2022wavlm}. While these methods have shown promise in controlled settings, they treat speech as a homogeneous signal and largely ignore its internal phonetic structure. This is a critical limitation, as emotional expression is not uniformly distributed over time but is realized through phoneme-level variations in duration, pitch (F0), and spectral characteristics \cite{schroder2001acoustic, barhate2016prosodic, rao2014focus}. Consequently, synthesis artifacts introduced during emotional voice conversion are also phoneme-dependent, with certain phoneme categories more affected than others. Utterance-level models aggregate over this fine-grained structure and may miss key discriminative cues. In addition, Fursule et al.~\cite{fursule2026gender} highlight that standard performance metrics can obscure gender-based disparities in audio deepfake detection, while Kshirsagar et al.~\cite{kshirsagar2026investigating} show that improvements in speech quality do not necessarily lead to better detection performance, revealing a trade-off between perceptual quality and robustness.

Despite growing interest in fine-grained speech analysis, phoneme-level 
investigation of emotionally manipulated synthetic speech remains 
largely unexplored. Prior phoneme-level studies, including Temmar et al.\ 
\cite{temmar2025phonetic} and Nallaguntla et al.\ \cite{nallaguntla2026phonemedf}, 
have examined general (non-emotional) TTS and voice conversion systems and 
established that complex phonemes such as diphthongs and fricatives exhibit 
greater divergence from real speech. However, these works do not address the 
emotionally conditioned setting, where the synthesis process must simultaneously 
transfer emotional prosody and preserve phonetic fidelity. It remains unclear 
how phoneme-level distributions shift under emotional manipulation, which 
phoneme categories are most sensitive to EVC artifacts, and whether 
distributional divergence at the phoneme-level predicts classification 
performance under matched emotional conditions. Addressing these questions is 
essential for building interpretable and robust detectors of emotional audio 
deepfakes.
The main 
contributions of this work are as follows:

\begin{enumerate}

    \item We propose a controlled phoneme-level framework for analyzing emotional 
    deepfakes, comparing real and EVC-generated speech under matched emotional 
    conditions using shared transcripts and phoneme-aligned representations.

    \item We explore the correlation between WavLM embeddings with symmetric KLD and an RBF-kernel SVM 
    across four emotions and two EVC systems.

    \item We release a curated dataset with aligned transcripts and phoneme-level 
    TextGrid annotations to support reproducibility and future research.

\end{enumerate}

The remainder of this paper is organized as follows. Section~\ref{sec:related} 
reviews related work. Section~\ref{sec:methodology} presents the proposed 
phoneme-level framework. Section~\ref{sec:results} provides the results and 
analysis. Finally, Section~\ref{sec:conclusion} concludes the paper.

\section{RELATED WORK}
\label{sec:related}
\subsection{Audio Deepfake Detection}

The rapid improvement in synthetic speech quality has spurred parallel advances 
in audio deepfake detection. Early detection systems relied on handcrafted 
spectral features such as MFCCs, LFCCs, and constant-Q cepstral coefficients, 
which were effective against early vocoders but struggled to generalize to 
modern neural systems \cite{yi2023audio, delgado2021asvspoof}. The ASVspoof 
challenge series \cite{delgado2021asvspoof} has been instrumental in 
benchmarking detection systems and promoting generalization research. More 
recently, SSL-based representations have emerged as powerful frontends for 
detection. Baevski et al.\ \cite{baevski2020wav2vec} introduced wav2vec 2.0, 
which learns contextualized speech representations from unlabeled audio through 
contrastive self-supervised objectives. Chen et al.\ \cite{chen2022wavlm} 
proposed WavLM, which augments masked speech prediction with a denoising 
pretraining objective and achieves state-of-the-art performance.
\subsection{Phoneme-Level Analysis of Synthetic Speech}

A smaller but growing body of work has examined synthetic speech at the phoneme-level to gain interpretable insights into where and how synthesis artifacts 
arise. Yang et al.\ \cite{yang2025forensic} compared segmental phonetic features 
with global audio-level representations for forensic deepfake audio detection, 
showing that phoneme-level features encode complementary discriminative 
information not captured by utterance-level models. Temmar et al.\ 
\cite{temmar2025phonetic} conducted phoneme- and word-level analysis of real and 
synthetic speech using HuBERT embeddings, finding that diphthongs and fricatives 
deviate most strongly from real speech. 
Building on this line of work, Nallaguntla et al.\ \cite{nallaguntla2026phonemedf} 
introduced PhonemeDF, a large-scale phoneme-annotated dataset for deepfake 
detection and naturalness evaluation, and demonstrated that complex phonemes 
such as diphthongs and fricatives exhibit consistently higher divergence from 
real speech, while simpler monophthongs such as /AH/ are synthesized more 
reliably. Collectively, these studies establish the value of phoneme-level 
analysis for understanding synthesis artifacts in general TTS and voice 
conversion settings.
\subsection{Neural Speech Synthesis and Voice Conversion}

The rapid progress of neural speech synthesis has produced systems capable of 
generating highly natural and expressive speech. Autoregressive models and 
flow-based architectures \cite{ren2019fastspeech, prenger2019waveglow} have 
enabled high-fidelity text-to-speech (TTS) synthesis, while generative 
adversarial network (GAN) based voice conversion systems \cite{kameoka2018stargan} 
and autoencoder-based approaches \cite{qian2019autovc} have extended these 
capabilities to voice style transfer. In the emotional domain, Zhou et al.\ 
proposed VAW-GAN-CWT \cite{zhou2020converting}, a speaker-independent EVC 
system that disentangles speaker identity from emotional expression using 
variational autoencoders and GAN training. The same group later introduced 
DeepEST \cite{zhou2021seen}, which extended EVC to unseen emotional styles using 
a newly released emotional speech dataset (ESD) \cite{zhou2022emotional}. While 
these systems produce expressive and perceptually convincing emotional speech, 
they also introduce synthesis artifacts that manifest differently across phoneme 
categories, motivating a fine-grained phoneme-level investigation.

\subsection{Emotional Deepfake Detection}

Despite progress in both emotional voice conversion and phoneme-level deepfake 
analysis, the detection of \textit{emotionally manipulated} synthetic speech 
remains underexplored. The EmoFake dataset \cite{zhao2024emofake} represents an 
important step toward this direction, providing real and EVC-generated speech 
across multiple emotions to support research on emotion fake audio detection. 
However, existing work has not yet systematically examined how phoneme-level 
distributions shift under emotional voice conversion, which phoneme categories 
are most sensitive to EVC-induced artifacts, or how distributional divergence at 
the phoneme-level relates to detectability under matched emotional conditions. 
This work directly addresses these open questions through a controlled 
phoneme-level analysis of emotionally converted speech, extending prior 
phoneme-level frameworks to the emotionally conditioned synthesis setting.
\section{Methodology}
\label{sec:methodology}
\begin{figure*}[t]
    \centering
    \includegraphics[width=0.8\textwidth, keepaspectratio]{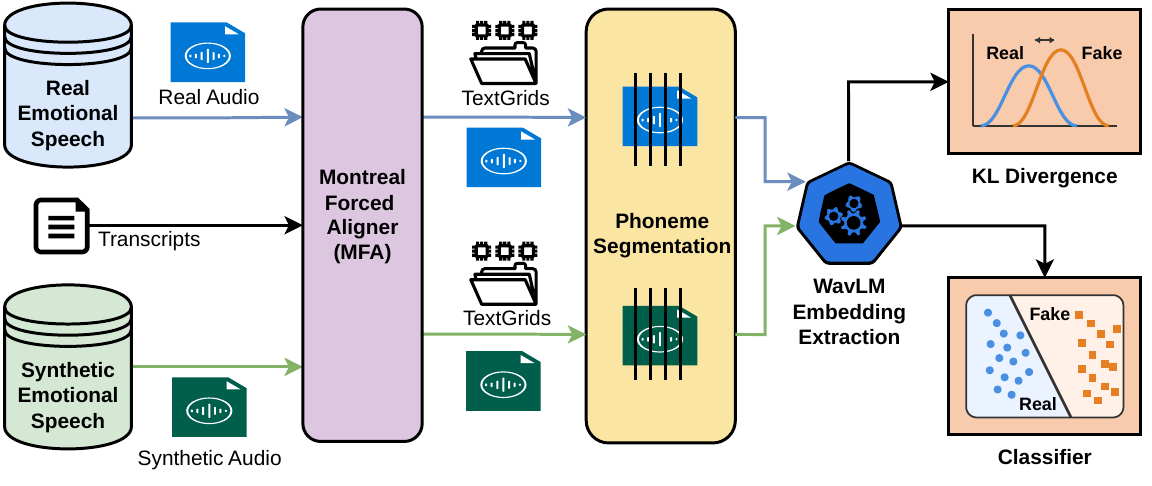}
    \caption{Overview of the proposed phoneme-level emotional deepfake analysis pipeline.}
    \label{fig:pipeline}
\end{figure*}

We propose a phoneme-level analysis framework for emotional deepfake detection, as illustrated in Fig.~\ref{fig:pipeline}. The framework operates on real and synthetic emotional speech in a controlled parallel setting. Synthetic emotional speech is generated from real neutral utterances using EVC models and compared against real speech of the same target emotion (e.g., synthetic angry vs.\ real angry). Since the linguistic content remains consistent across emotional conditions for the selected speakers, any observed differences arise from emotional transformation and synthesis effects rather than content variation. Audio–transcript pairs are processed using the Montreal Forced Aligner (MFA) \cite{mcauliffe2017montreal} to obtain time-aligned phoneme boundaries as TextGrid files. These alignments segment each signal into phoneme-level units. For each segment, WavLM embeddings are extracted to construct phoneme-level representations. Distributional differences between real and synthetic speech are quantified using symmetric Kullback–Leibler divergence (KLD), and phoneme-wise classification is performed using a Support Vector Machine (SVM) with a radial basis function (RBF) kernel.

In this section, we describe the components of the proposed framework in detail, including the dataset, phoneme alignment, feature extraction, classification, and evaluation methodology.
\subsection{Dataset}

We use the EmoFake dataset \cite{zhao2024emofake}, which contains both real and synthetic emotional speech. The real audio is derived from the Emotional Speech Dataset (ESD) \cite{zhou2022emotional}. We construct a controlled dataset consisting of real and synthetic speech across four target emotions: Angry, Happy, Sad, and Surprise. For the selected speakers, the same set of utterances is used across all emotional conditions, ensuring identical linguistic content and enabling aligned phoneme-level comparison without transcription mismatch.

We restrict our analysis to speakers whose emotional speech is generated from real neutral utterances and for whom all target emotions are available. In the EmoFake dataset, only speakers 0011 and 0016 satisfy this condition, as other speakers either lack complete emotion coverage or use different source emotions. Their speech has been converted into all target emotions using two emotional voice conversion (EVC) models: VAW-GAN-CWT (EVC1) \cite{zhou2020converting} and DeepEST (EVC2) \cite{zhou2021seen}. All audio signals are resampled to 16~kHz. The final dataset consists of 700 real utterances per emotion and 700 synthetic utterances per system per emotion, yielding 5600 synthetic emotional samples in total.

\subsection{Phoneme-Level Alignment and Segmentation}

To obtain fine-grained phoneme boundaries, we employ MFA \cite{mcauliffe2017montreal} with the \textit{english\_us\_arpa.dict} pronunciation dictionary and the \textit{english\_us\_arpa.zip} acoustic model. Each audio signal is aligned to an ARPAbet phoneme sequence using its corresponding transcript, producing time-aligned TextGrid annotations. Speech signals are then segmented into phoneme-level units, where each segment is treated as an independent sample. Stress markers are removed from phoneme labels (e.g., AA0, AA1, AA2~$\rightarrow$~AA) to ensure label consistency. Phonemes with insufficient samples are excluded to ensure reliable statistical estimation; for example, /ZH/ is excluded due to only 4 occurrences.

\subsection{Feature Extraction}

Phoneme-level representations are extracted using the pretrained WavLM\footnote{\url{https://huggingface.co/microsoft/wavlm-large}} \cite{chen2022wavlm} self-supervised speech model. For each phoneme segment, frame-level embeddings are obtained from the final hidden layer of the model. A fixed-dimensional phoneme-level representation is derived by mean pooling over the $T$ frames within each segment:

\begin{equation}
\mathbf{z} = \frac{1}{T} \sum_{t=1}^{T} \mathbf{h}_t
\end{equation}

where $\mathbf{z}$ denotes the phoneme-level embedding and $\mathbf{h}_t$ is the frame-level hidden state at time $t$.

\subsection{Classifier }

To evaluate the discriminability between real and synthetic phoneme representations, we perform phoneme-level binary classification using an SVM with an RBF kernel \cite{cortes1995support}. For each phoneme, embeddings from real and synthetic speech of the same target emotion are treated as separate classes, resulting in 38 independent binary classifiers. The RBF kernel enables the model to capture nonlinear separability between real and synthetic distributions.
\subsection{ Evaluation Metrics}
We employed the following two evaluation metrics for performance assessment
\subsubsection{Accuracy}
Classification performance is evaluated using accuracy, which reflects the degree of separability between real and synthetic phoneme distributions. 
\begin{equation}
\text{Accuracy} = \frac{TP + TN}{TP + TN + FP + FN}
\end{equation}

where $TP$ and $TN$ denote the number of true positives and true negatives, and $FP$ and $FN$ represent false positives and false negatives, respectively.

\subsubsection{Kullback-Leibler divergence (KLD)}

To quantify distributional differences between real and synthetic phoneme representations under matched emotional conditions, we compute the symmetric Kullback-Leibler divergence (KLD). For each phoneme, embeddings from real and synthetic speech are modeled as multivariate Gaussian distributions $\mathcal{N}_R(\mu_R, \Sigma_R)$ and $\mathcal{N}_S(\mu_S, \Sigma_S)$, where $\mu$ is the mean vector and $\Sigma$ the covariance matrix. The symmetric KLD is defined as:

\begin{equation}
D_{\mathrm{sym}} = \frac{1}{2} \left[ D_{\mathrm{KL}}(\mathcal{N}_R \| \mathcal{N}_S) + D_{\mathrm{KL}}(\mathcal{N}_S \| \mathcal{N}_R) \right]
\label{eq:symkld}
\end{equation}

Higher $D_{\mathrm{sym}}$ values indicate larger distributional differences between real and synthetic phoneme embeddings, capturing deviations introduced by the synthesis process. We hypothesize that phonemes with higher divergence are more sensitive to synthesis artifacts and will exhibit higher classification accuracy.
\subsubsection{Pearson correlation coefficient}
To assess the relationship between distributional divergence and detectability, we compute the Pearson correlation coefficient $r$ between phoneme-level KLD values and classification accuracies across all phonemes within each condition. The null hypothesis assumes no linear relationship between the two variables, $H_0: r = 0$.

The Pearson correlation coefficient is defined as:
\begin{equation}
r = \frac{\sum_{i=1}^{N} (x_i - \mu_x)(y_i - \mu_y)}
{\sqrt{\sum_{i=1}^{N} (x_i - \mu_x)^2} \sqrt{\sum_{i=1}^{N} (y_i - \mu_y)^2}}
\end{equation}

where $x_i$ and $y_i$ denote the KLD value and classification accuracy for the $i$-th phoneme, respectively, $\mu_x$ and $\mu_y$ are their corresponding means, and $N$ is the total number of phonemes.

\section{Results \& Discussion}
\label{sec:results}
\begin{table*}[t]
\centering
\caption{Vowel phonemes: KLD and RBF-SVM accuracy (\%) comparing real vs.\ fake emotional speech}
\label{tab:vowel_target}
\resizebox{\textwidth}{!}{%
\begin{tabular}{l cc cc cc cc cc cc cc cc}
\toprule
 & \multicolumn{2}{c}{\textbf{EVC1-Angry}} & \multicolumn{2}{c}{\textbf{EVC1-Happy}} & \multicolumn{2}{c}{\textbf{EVC1-Sad}} & \multicolumn{2}{c}{\textbf{EVC1-Surprise}} & \multicolumn{2}{c}{\textbf{EVC2-Angry}} & \multicolumn{2}{c}{\textbf{EVC2-Happy}} & \multicolumn{2}{c}{\textbf{EVC2-Sad}} & \multicolumn{2}{c}{\textbf{EVC2-Surprise}} \\
\cmidrule(lr){2-3}\cmidrule(lr){4-5}\cmidrule(lr){6-7}\cmidrule(lr){8-9}\cmidrule(lr){10-11}\cmidrule(lr){12-13}\cmidrule(lr){14-15}\cmidrule(lr){16-17}
\textbf{Phoneme} & \textbf{KLD} & \textbf{Acc} & \textbf{KLD} & \textbf{Acc} & \textbf{KLD} & \textbf{Acc} & \textbf{KLD} & \textbf{Acc} & \textbf{KLD} & \textbf{Acc} & \textbf{KLD} & \textbf{Acc} & \textbf{KLD} & \textbf{Acc} & \textbf{KLD} & \textbf{Acc} \\
\midrule
AA & 15.95 & 80.1 & 15.66 & 81.6 & 13.89 & 81.9 & 16.81 & 78.1 & 15.06 & 82.4 & 14.41 & 76.6 & 15.05 & 79.0 & 12.92 & 70.1 \\
AE & 14.62 & 83.8 & 13.82 & 75.3 & 13.35 & 85.6 & 13.78 & 82.1 & 11.29 & 78.4 & 11.96 & 70.9 & 12.50 & 79.8 & 11.45 & 73.2 \\
AH & 8.64 & 77.8 & 7.98 & 80.0 & 9.18 & 77.6 & 8.07 & 79.9 & 7.32 & 72.3 & 6.30 & 70.6 & 7.99 & 77.6 & 6.04 & 72.8 \\
AO & 29.50 & 87.5 & 31.83 & 83.3 & 33.86 & 84.9 & 43.41 & 79.6 & 26.06 & 82.7 & 29.26 & 79.2 & 29.96 & 73.1 & 26.31 & 73.5 \\
AW & 29.16 & 87.3 & 24.65 & 83.3 & 21.79 & 79.6 & 27.63 & 83.3 & 22.22 & 85.2 & 24.31 & 90.7 & 20.86 & 85.2 & 26.57 & 74.1 \\
AY & 19.65 & 85.4 & 20.96 & 84.8 & 22.51 & 86.5 & 19.54 & 89.0 & 12.78 & 78.7 & 16.59 & 76.2 & 18.61 & 88.3 & 14.41 & 74.4 \\
EH & 11.57 & 77.5 & 11.67 & 75.0 & 14.73 & 86.7 & 9.82 & 76.0 & 9.30 & 77.6 & 9.66 & 68.9 & 12.20 & 79.3 & 9.21 & 72.7 \\
ER & 19.22 & 81.6 & 20.30 & 76.9 & 18.11 & 80.9 & 21.31 & 79.4 & 16.91 & 80.7 & 16.76 & 73.5 & 17.50 & 86.8 & 14.35 & 71.5 \\
EY & 19.28 & 90.1 & 20.16 & 83.5 & 25.26 & 89.8 & 18.96 & 87.0 & 16.18 & 80.2 & 17.26 & 81.7 & 21.31 & 76.9 & 18.39 & 76.9 \\
IH & 11.21 & 80.8 & 8.53 & 72.2 & 11.73 & 78.7 & 8.18 & 82.2 & 7.84 & 71.6 & 6.93 & 70.1 & 9.20 & 80.4 & 6.59 & 69.7 \\
IY & 13.42 & 82.8 & 11.01 & 73.0 & 13.35 & 86.7 & 10.96 & 81.2 & 11.23 & 73.0 & 10.69 & 76.6 & 12.97 & 78.3 & 10.11 & 74.8 \\
OW & 23.59 & 86.2 & 23.23 & 80.2 & 25.09 & 87.4 & 23.08 & 80.2 & 17.03 & 88.5 & 20.59 & 80.2 & 20.88 & 89.7 & 17.89 & 75.6 \\
OY & 22.14 & 75.0 & 25.04 & 75.0 & 23.82 & 87.5 & 34.77 & 75.0 & 19.50 & 81.2 & 20.96 & 93.8 & 23.90 & 93.8 & 22.27 & 75.0 \\
UH & 53.21 & 83.8 & 51.30 & 92.1 & 40.53 & 86.5 & 64.29 & 89.7 & 40.99 & 83.8 & 36.58 & 74.4 & 34.48 & 83.8 & 42.86 & 78.9 \\
UW & 21.01 & 83.3 & 18.88 & 84.1 & 20.90 & 89.2 & 22.13 & 87.0 & 19.26 & 86.1 & 17.91 & 86.9 & 18.67 & 92.8 & 19.29 & 76.9 \\
\bottomrule
\end{tabular}}
\end{table*}
 
\begin{table*}[t]
\centering
\caption{Consonant phonemes: KLD and RBF-SVM accuracy (\%) comparing real vs.\ fake emotional speech}
\label{tab:consonant_target}
\resizebox{\textwidth}{!}{%
\begin{tabular}{l cc cc cc cc cc cc cc cc}
\toprule
 & \multicolumn{2}{c}{\textbf{EVC1-Angry}} & \multicolumn{2}{c}{\textbf{EVC1-Happy}} & \multicolumn{2}{c}{\textbf{EVC1-Sad}} & \multicolumn{2}{c}{\textbf{EVC1-Surprise}} & \multicolumn{2}{c}{\textbf{EVC2-Angry}} & \multicolumn{2}{c}{\textbf{EVC2-Happy}} & \multicolumn{2}{c}{\textbf{EVC2-Sad}} & \multicolumn{2}{c}{\textbf{EVC2-Surprise}} \\
\cmidrule(lr){2-3}\cmidrule(lr){4-5}\cmidrule(lr){6-7}\cmidrule(lr){8-9}\cmidrule(lr){10-11}\cmidrule(lr){12-13}\cmidrule(lr){14-15}\cmidrule(lr){16-17}
\textbf{Phoneme} & \textbf{KLD} & \textbf{Acc} & \textbf{KLD} & \textbf{Acc} & \textbf{KLD} & \textbf{Acc} & \textbf{KLD} & \textbf{Acc} & \textbf{KLD} & \textbf{Acc} & \textbf{KLD} & \textbf{Acc} & \textbf{KLD} & \textbf{Acc} & \textbf{KLD} & \textbf{Acc} \\
\midrule
B  & 26.74 & 78.9 & 26.32 & 69.5 & 34.16 & 87.4 & 25.64 & 83.2 & 28.52 & 69.5 & 28.63 & 84.2 & 34.51 & 83.2 & 26.59 & 76.8 \\
CH & 48.79 & 76.2 & 47.15 & 83.7 & 33.32 & 83.7 & 53.37 & 88.4 & 38.76 & 88.1 & 42.85 & 86.0 & 30.95 & 93.0 & 51.24 & 86.0 \\
D  & 11.80 & 78.4 & 10.07 & 75.7 & 15.05 & 81.8 & 11.25 & 78.9 & 10.67 & 73.3 & 12.34 & 67.8 & 13.81 & 76.6 & 14.49 & 77.2 \\
DH & 22.53 & 78.6 & 18.16 & 75.7 & 31.87 & 75.7 & 19.31 & 72.2 & 19.65 & 81.4 & 19.23 & 77.8 & 31.19 & 80.6 & 20.70 & 73.6 \\
F  & 18.47 & 84.9 & 18.34 & 77.4 & 14.16 & 83.0 & 20.12 & 81.1 & 16.07 & 81.1 & 18.72 & 76.4 & 14.75 & 67.0 & 18.04 & 79.2 \\
G  & 16.82 & 87.1 & 21.17 & 85.7 & 26.32 & 85.7 & 21.17 & 80.0 & 19.04 & 72.9 & 18.46 & 81.4 & 23.23 & 80.0 & 22.93 & 78.6 \\
HH & 22.88 & 75.4 & 18.32 & 76.9 & 18.78 & 83.1 & 17.18 & 77.3 & 18.18 & 68.5 & 19.24 & 75.4 & 17.31 & 70.0 & 18.43 & 77.3 \\
JH & 51.58 & 84.4 & 45.51 & 86.7 & 37.51 & 88.9 & 49.29 & 88.9 & 44.93 & 86.7 & 45.67 & 82.2 & 37.35 & 88.9 & 42.62 & 86.7 \\
K  & 11.13 & 82.5 & 12.60 & 75.8 & 16.24 & 82.6 & 11.55 & 76.9 & 9.98  & 81.9 & 11.48 & 72.7 & 14.91 & 82.0 & 10.94 & 76.9 \\
L  & 11.50 & 74.3 & 10.99 & 74.8 & 11.56 & 73.6 & 10.70 & 74.8 & 11.09 & 70.8 & 11.07 & 73.8 & 11.59 & 79.1 & 8.54  & 71.8 \\
M  & 10.31 & 73.0 & 10.50 & 65.5 & 13.43 & 84.5 & 10.62 & 75.9 & 10.40 & 77.0 & 11.47 & 79.3 & 12.12 & 80.5 & 8.32  & 73.0 \\
N  & 7.82  & 77.0 & 7.77  & 79.8 & 10.47 & 82.5 & 7.69  & 81.3 & 6.24  & 73.7 & 6.98  & 75.7 & 8.66  & 79.8 & 6.14  & 74.9 \\
NG & 25.04 & 85.4 & 22.16 & 87.5 & 22.02 & 79.2 & 23.86 & 87.5 & 22.52 & 85.4 & 22.22 & 85.4 & 25.86 & 83.3 & 26.11 & 83.3 \\
P  & 18.21 & 78.2 & 16.86 & 74.5 & 20.16 & 84.5 & 17.19 & 80.9 & 14.50 & 71.8 & 14.49 & 81.8 & 18.32 & 85.5 & 13.63 & 71.8 \\
R  & 10.48 & 75.2 & 14.12 & 85.3 & 12.41 & 79.2 & 14.73 & 79.5 & 11.29 & 72.2 & 11.54 & 74.2 & 12.37 & 77.9 & 9.28  & 70.9 \\
S  & 13.09 & 76.1 & 13.36 & 81.6 & 9.62  & 77.9 & 15.59 & 79.1 & 12.05 & 82.9 & 13.87 & 79.1 & 10.92 & 79.6 & 14.33 & 77.4 \\
SH & 48.56 & 78.3 & 42.07 & 85.0 & 28.71 & 65.0 & 48.39 & 85.0 & 37.60 & 71.7 & 43.61 & 85.0 & 27.97 & 70.0 & 40.29 & 75.0 \\
T  & 7.62  & 74.0 & 7.35  & 73.6 & 11.47 & 76.2 & 8.42  & 75.4 & 6.78  & 69.7 & 6.68  & 71.3 & 11.26 & 76.5 & 7.70  & 77.7 \\
TH & 32.04 & 74.0 & 31.32 & 90.4 & 25.05 & 86.8 & 28.17 & 88.5 & 27.26 & 72.0 & 32.30 & 86.8 & 26.40 & 82.7 & 24.93 & 84.6 \\
V  & 24.48 & 81.7 & 22.29 & 73.2 & 30.83 & 81.7 & 22.98 & 81.7 & 20.39 & 86.6 & 19.07 & 85.4 & 24.88 & 89.0 & 18.55 & 82.9 \\
W  & 18.55 & 79.6 & 22.34 & 76.6 & 17.71 & 79.6 & 20.14 & 79.6 & 17.95 & 75.2 & 21.81 & 78.8 & 16.23 & 75.2 & 15.93 & 78.8 \\
Y  & 28.28 & 75.7 & 30.19 & 76.8 & 29.48 & 95.7 & 28.04 & 74.3 & 29.95 & 78.6 & 28.72 & 72.5 & 29.85 & 75.4 & 26.90 & 71.4 \\
Z  & 15.38 & 83.1 & 14.18 & 84.9 & 13.53 & 81.9 & 15.68 & 78.9 & 14.56 & 78.3 & 14.58 & 78.3 & 11.99 & 81.9 & 15.75 & 74.7 \\
\bottomrule
\end{tabular}}
\end{table*}

In this section, we discuss our results in the context of existing literature.

\subsection{Phoneme-Level Analysis}

Tables~\ref{tab:vowel_target} and~\ref{tab:consonant_target} present 
the phoneme-level KLD values and RBF-SVM classification accuracies for 
vowel and consonant categories across different emotional conditions and 
synthesis systems. The results reveal consistent phoneme-dependent 
patterns that align with and extend prior findings in the deepfake 
detection literature.

For vowel phonemes, complex vowels and diphthongs such as /AO/, /AW/, 
/AY/, and /OY/ consistently exhibit higher KLD values across both EVC1 
and EVC2 systems, indicating larger distributional deviations between 
real and synthetic speech. In particular, /UH/ reaches the largest KLD 
values overall (e.g., 64.29 under EVC1-Surprise), and /AO/ shows 
consistently high divergence across all four emotions. This pattern is 
consistent with the findings of Temmar et al.\ \cite{temmar2025phonetic}, 
who similarly observed that diphthongs exhibit the strongest deviations 
from real speech. 
Our results extend this finding to the emotionally conditioned synthesis 
setting, suggesting that the acoustic complexity of diphthongs and tense 
vowels makes them particularly difficult to synthesize faithfully 
regardless of whether the conversion target is a neutral or emotional 
style. Phonemes with higher KLD values achieve relatively high 
classification accuracies under several conditions, supporting the 
hypothesis that larger distributional differences correspond to improved 
separability under nonlinear modeling 
\cite{nallaguntla2026phonemedf}. Vowels such as /AY/, /EY/, and 
/OW/ demonstrate stable and consistently high classification accuracies 
across emotional conditions despite moderate KLD values, suggesting that 
the RBF-SVM is able to exploit nonlinear structure in the embedding 
space beyond what divergence alone captures. Simpler monophthongs, 
including /AH/ and /IH/, tend to exhibit the lowest KLD values and 
moderate but stable classification performance across both EVC systems.

For consonant phonemes, a similar but distinct pattern emerges. 
Fricatives and affricates such as /CH/, /JH/, and /SH/ exhibit the 
highest KLD values among consonants, reflecting their well-known 
sensitivity to spectral modeling, a finding that corroborates Temmar 
et al. \cite{temmar2025phonetic}, who identified fricatives as 
providing among the most interpretable phoneme-level cues for deepfake 
detection. The high spectral complexity and aperiodic energy 
distribution of fricatives make them inherently difficult to reproduce 
under voice conversion, and our results indicate that this challenge 
persists when emotional prosody is additionally imposed during 
conversion. In contrast, plosives such as /T/, /K/, and /P/ show 
lower KLD values but maintain stable classification accuracy across 
conditions, suggesting that their simpler and more transient temporal 
structures are preserved more reliably during emotional conversion. 
 Nasals and approximants, including 
/M/, /N/, /L/, and /R/, generally exhibit lower divergence and moderate 
classification performance across both systems and emotions, indicating 
that these sonorant categories are comparatively robust to synthesis 
artifacts. Yang et al.\ \cite{yang2025forensic} similarly noted that 
sonorant-like segments tend to be less discriminative in segmental 
deepfake detection, as their smooth spectral transitions are more 
consistently captured by generative models.

Overall, these results demonstrate that phoneme-level behavior 
varies significantly across phoneme classes even under matched emotional 
conditions, and that this variation follows principled acoustic 
distinctions. Complex vowels and fricatives  characterized by greater 
spectral detail, dynamic formant trajectories, or aperiodic energy — 
remain highly sensitive to EVC artifacts and tend to be the most 
discriminative for detection. Simpler phonemes with more stable spectral 
profiles are more consistently reproduced and less useful as individual 
discriminative cues. Crucially, this hierarchy of phoneme sensitivity 
observed in the emotional synthesis setting mirrors what has been 
reported for general (non-emotional) TTS systems 
\cite{temmar2025phonetic, nallaguntla2026phonemedf}, suggesting that 
phoneme-dependent synthesis difficulty is a general property of 
neural voice conversion rather than a consequence of the specific 
emotional manipulation applied.

\begin{figure}[t]
    \centering
    \includegraphics[width=\linewidth]{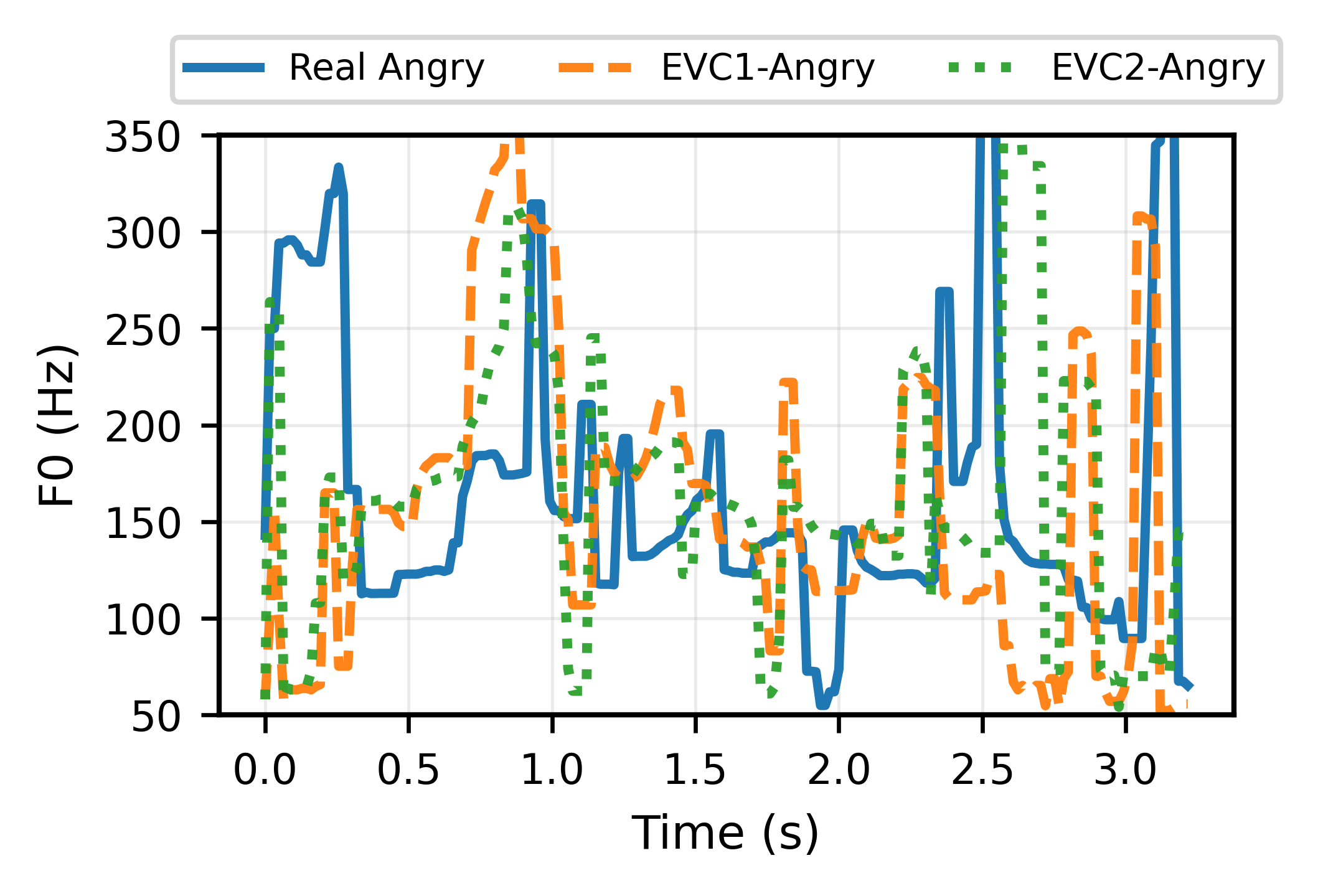}
    \caption{F0 contour comparison for a representative utterance between real and synthetic angry speech (EVC1, EVC2) generated from neutral speech.}
    \label{fig:f0_example}
\end{figure}

Figure~\ref{fig:f0_example} illustrates the F0 contours for a 
representative real angry speech utterance and its synthetic 
counterparts generated by EVC1 and EVC2 from real neutral speech. 
While both systems broadly capture the elevated pitch range 
characteristic of angry speech, the synthetic signals exhibit 
markedly irregular transitions and abrupt variations compared to 
the smoother, temporally coherent F0 trajectory of real speech, consistent with the established understanding that natural emotional 
expression requires coordinated prosodic gestures across time 
\cite{schroder2001acoustic, barhate2016prosodic}. The two systems produce distinct F0 
artifact profiles, with EVC1 showing sharper pitch discontinuities 
and EVC2 exhibiting somewhat smoother but still irregular dynamics, 
reflecting the architecture specific artifact signatures reported 
in the broader deepfake detection literature \cite{yi2023audio} and 
mirroring the system-dependent KLD differences observed in 
Tables~\ref{tab:vowel_target} and~\ref{tab:consonant_target}. 
These prosodic irregularities complement the phoneme-level 
distributional analysis, supporting the observation of Yang et al.\ 
\cite{yang2025forensic} that prosodic and segmental features carry 
complementary discriminative information for deepfake detection.

\subsection{Correlation Analysis}

Table~\ref{tab:corr_combined}  presents the Pearson correlation 
between phoneme-level KLD and RBF-SVM classification accuracy 
across all conditions. A generally positive relationship is observed 
for both vowel and consonant phonemes, with several strong and 
statistically significant correlations, notably EVC1-Happy 
($r=0.75$, $p=0.0012$) for vowels and EVC1-Surprise 
($r=0.69$, $p=0.0002$) for consonants indicating that 
phonemes with higher distributional divergence tend to be more easily 
separable under nonlinear decision boundaries, consistent with 
findings reported by Temmar et al.\ \cite{temmar2025phonetic} and 
Nallaguntla et al.\ \cite{nallaguntla2026phonemedf}. Weaker 
correlations in certain conditions, such as EVC2-Sad ($r=0.19$, 
$p=0.4900$), suggest that low-arousal emotions introduce 
less prosodic variability and reduce the alignment between divergence 
and detectability \cite{schroder2001acoustic, barhate2016prosodic}. Overall, KLD serves 
as a meaningful predictor of discriminability when paired with 
nonlinear classifiers, highlighting the importance of jointly 
considering statistical divergence and model capacity for 
phoneme-level emotional deepfake analysis.

\begin{table}[t]
\centering
\small
\setlength{\tabcolsep}{8pt}
\renewcommand{\arraystretch}{1.1}

\caption{Pearson correlation between KLD and RBF-SVM accuracy for vowel and consonant phonemes across all conditions.}
\label{tab:corr_combined}

\begin{tabular}{lcc|cc}
\toprule
 & \multicolumn{2}{c}{\textbf{Vowels}} & \multicolumn{2}{c}{\textbf{Consonants}} \\
\cmidrule(lr){2-3} \cmidrule(lr){4-5}
\textbf{Condition} & $r$ & $p$-value & $r$ & $p$-value \\
\midrule
EVC1-Angry    & $0.37$ & $0.1770$ & $0.16$ & $0.4670$ \\
EVC2-Angry    & $0.63$ & $0.0115$ & $0.35$ & $0.1060$ \\
EVC1-Happy    & $0.75$ & $0.0012$ & $0.46$ & $0.0279$ \\
EVC2-Happy    & $0.45$ & $0.0947$ & $0.63$ & $0.0013$ \\
EVC1-Sad      & $0.46$ & $0.0880$ & $0.29$ & $0.1860$ \\
EVC2-Sad      & $0.19$ & $0.4900$ & $0.39$ & $0.0674$ \\
EVC1-Surprise & $0.31$ & $0.2680$ & $0.69$ & $0.0002$ \\
EVC2-Surprise & $0.68$ & $0.0053$ & $0.59$ & $0.0030$ \\
\bottomrule
\end{tabular}
\end{table}

\section{CONCLUSION}
\label{sec:conclusion}

Detecting emotionally manipulated synthetic speech remains 
challenging, as modern EVC systems closely mimic the acoustic 
and prosodic characteristics of natural emotional expression 
while existing utterance-level approaches overlook fine-grained 
phonetic structure. In this work, we proposed a phoneme-level 
framework to address this gap. Our results show that 
complex vowels and fricatives exhibit the highest distributional 
divergence and classification accuracy, while simpler phonemes 
remain stable under emotional voice conversion. A generally 
positive correlation between KLD and classification performance 
confirms that distributional divergence is a meaningful 
predictor of detectability. Future work will extend 
this framework to emotion-to-emotion conversion models, 
additional languages, and multimodal prosodic features to 
improve generalization and robustness.


\addtolength{\textheight}{-12cm}   




\end{document}